\documentclass{vgtc}                          

\ifpdf
  \pdfoutput=1\relax                   
  \usepackage{graphicx}                
  \DeclareGraphicsExtensions{.pdf,.png,.jpg,.jpeg} 
\else
  \usepackage{graphicx}                
  \DeclareGraphicsExtensions{.eps}     
\fi%

\graphicspath{{figures/}{pictures/}{images/}{./}} 

\usepackage{microtype}                 
\PassOptionsToPackage{warn}{textcomp}  
\usepackage{textcomp}                  
\usepackage{mathptmx}                  
\usepackage{times}                     
\usepackage{cite}                      
\usepackage{tabu}                      
\usepackage{booktabs}                  

\onlineid{0}

\vgtccategory{Research}

\vgtcinsertpkg

\title{Standing Balance Improvement Using Vibrotactile Feedback in Virtual Reality}

\author{M. Rasel Mahmud\thanks{e-mail: m.raselmahmud1@gmail.com}\\  %
     \parbox{1.4in}{\scriptsize \centering Computer Science \\ The University of Texas at San Antonio} %
\and Michael Stewart\thanks{e-mail: michael.stewart@utsa.edu }\\ %
     \parbox{1.4in}{\scriptsize \centering Kinesiology \\ The University of Texas at San Antonio} %
\and Alberto Cordova\thanks{e-mail: Alberto.Cordova@utsa.edu}\\ %
     \parbox{1.4in}{\scriptsize \centering Kinesiology \\ The University of Texas at San Antonio} %
\and John Quarles\thanks{e-mail: John.Quarles@utsa.edu}\\ %
     \parbox{1.4in}{\scriptsize \centering Computer Science \\ The University of Texas at San Antonio}}

\abstract{
Virtual Reality (VR) users often encounter postural instability, i.e., balance issues, which can be a significant impediment to universal usability and accessibility, particularly for those with balance impairments. Prior research has validated imbalance issues, but little effort has been made to mitigate them. We recruited 39 participants (with balance impairments: 18, without balance impairments: 21) to examine the effect of various vibrotactile feedback techniques on balance in virtual reality, specifically spatial vibrotactile, static vibrotactile, rhythmic vibrotactile, and vibrotactile feedback mapped to the center of pressure (CoP). Participants completed standing visual exploration and standing reach and grasp tasks. According to within-subject results, each vibrotactile feedback enhanced balance in VR significantly (\textit{p} $<$ .001) for those with and without balance impairments. Spatial and CoP vibrotactile feedback enhanced balance significantly more (\textit{p} $<$ .001) than other vibrotactile feedback. This study presents strategies that might be used in future virtual environments to enhance standing balance and bring VR closer to universal usage.

} 

\CCScatlist{
  \CCScatTwelve{Human-centered computing}{Visu\-al\-iza\-tion}{Visu\-al\-iza\-tion techniques}{Treemaps};
  \CCScatTwelve{Human-centered computing}{Visu\-al\-iza\-tion}{Visualization design and evaluation methods}{}
}

\begin{document}


\firstsection{Introduction}

\maketitle

Over one billion individuals, or 15\% of the global population, have a disability \cite{WinNT}. Virtual reality (VR) is inaccessible to a significant number of individuals with disabilities \cite{agrawal2009disorders,ferdous2016visual,ferdous2018investigating,guo2013effects,samaraweera2013latency}. Unfortunately, these people are seldomly considered throughout VR research and development, resulting in experiences that are exclusive and inaccessible. For instance, individuals with balance impairments (BI) may not be able to stand during VR encounters comfortably. This restriction may prohibit consumers from partaking in some portions of the virtual reality experience. VR Head-Mounted Displays (HMDs) severely destabilize users with BI as well as those without BI \cite{epure2014effect,lott2003effect,robert2016effect}. Nevertheless, minimal research has been performed to mitigate this effect. If these disparities could be rectified, people with and without BI might benefit from consumer VR technologies more easily.

In the field of assistive technology research, multiple feedback approaches of different modalities \cite{franco2012ibalance,sienko2017role} have been developed to enhance real-world balance. For instance, researchers have employed vibrotactile feedback to enhance the balance of individuals with impaired eyesight \cite{velazquez2010wearable}. Also, Mahmud et al. \cite{mahmud2022vibrotactile} investigated the effect of vibrotactile feedback in virtual walking. The influence of auditory feedback on gait in VR was also investigated in \cite{WinNT9}. However, very few studies have been conducted on how vibrotactile feedback affects balance in VR.


In this study, participants (i.e., persons with and without BI) attempted to maintain balance while standing in virtual environments (VE) with different methods of vibrotactile feedback (i.e., spatial, static, rhythmic, and vibrotactile feedback based on the center of pressure (CoP)). This research aims to make immersive VR more accessible to all individuals. The findings are intended to provide future VR developers with an understanding of how vibrotactile feedback may be used to enhance VR accessibility. Our proposed contributions include the following:
\begin{itemize}
\item We developed four novel vibrotactile feedback (spatial, static, rhythmic, CoP) techniques for balance improvement in VR.
\item  We conducted a study to determine the effects of our vibrotactile feedback techniques on standing balance in VR.
\item To improve generalizability, we recruited participants with balance impairments due to Multiple Sclerosis (MS), who are rarely considered in VR, and participants without MS.
\end{itemize}

\section{Background}
\subsection{Imbalance in VR}
Although VR systems have great applicability and have increased in utilization over the years, these systems possess several limiting factors that make them less accessible and usable for all, especially for populations with balance impairments (e.g., persons with multiple sclerosis (MS)). One of the primary limitations reported in the previous literature is that VR experiences can potentially negatively influence an individual’s postural control mechanisms \cite{takahashi2001change}. Postural control and balance are maintained using a variety of sensory feedback information regulated by the central nervous system \cite{diener1988role}. Sensory feedback information is predominately acquired independently from visual, vestibular, and proprioceptive systems to maintain stability and balance \cite{denomme2014understanding}. Aging and neurological diseases (e.g., MS) often contribute to the impairment of sensory feedback systems which can subsequently deteriorate balance and stability in these individuals \cite{borah2007age, kido2010postural, blaszczyk2007assessment, bloem1992postural}. 

These feedback information systems are particularly important in VR experiences that utilize HMDs. Particularly, VR HMDs can disrupt visual feedback and manipulate an individual’s proprioceptive feedback systems due to end-to-end latency \cite{martinez2018analysing, soltani2020influence}. Moreover, balance in immersive VR can be impaired due to the dissimilarity of tracking space between the virtual environment and the perceived physical stimulus \cite{niehorster2017accuracy, martinez2018analysing}. Although balance loss in HMD-based VR experiences has been acknowledged, minimal efforts have been made to mitigate these issues.

There is considerable research that has investigated the application of VR in rehabilitative settings aimed at improving postural control and gait performance in individuals \cite{bergeron2015use, cho2016treadmill, bisson2007functional, de2016effect, duque2013effects, meldrum2015effectiveness, park2015effects, rendon2012effect}. However, the majority of prior work has often been conducted in non-immersive VR environments without the use of HMDs. Moreover, most of the previous research on assistive feedback in VR has been primarily focused on the application of visual feedback \cite{ferdous2018investigating}. Therefore, to address these matters, this study investigates the application of assistive technology in the form of vibrotactile feedback to improve balance in immersive VR.

\subsection{Vibrotactile Feedback for Balance Improvement in the Real-World }
Previous research has suggested that providing different inputs such as visual \cite{alahakone2010real}, auditory \cite{chiari2005audio}, or vibrotactile \cite{sienko2012biofeedback} feedback in real-world applications can help mitigate instability issues and excessive postural sway experienced by individuals with balance impairments \cite{henry2019age, wannstedt1978use, sienko2018potential}. However, the application of vibrotactile feedback is often preferred over other assistive feedback modalities because vibrotactile feedback is thought to interfere minimally with other senses like seeing or hearing, which can be inhibited by visual or auditory feedback systems \cite{goodworth2009influence, wall2009vibrotactile}. The use of vibrotactile feedback to improve balance in real-world applications, such as in rehabilitative scenarios has been investigated in various previous studies. 

For example, Kingma et al. recruited 39 participants with an imbalance from severe bilateral vestibular loss to investigate how vibrotactile feedback affects balance and mobility in the real world \cite{kingma2019vibrotactile}. Participants wore a tactor (i.e., vibrotactile motor that makes vibration) belt around the waist for two hours every day for one month. If they were moving, they were required to wear the belt. The 12 tactors in the belt were activated via a microprocessor. They asked participants verbally to rate balance and mobility scores on a scale of 0 to 10 before and after one month of daily use of the belt. The average mobility and balance scores increased significantly (\textit{p} $<$.00001) compared to without the belt.

Rust et al. investigated the effect of vibrotactile feedback on trunk sway for fifteen participants with MS in the real world \cite{rust2020benefits}. Participants wore a headband with eight 150 Hz vibrators which were positioned at 45-degree intervals. Vibrators were activated when a sway threshold in the vibrator's direction was surpassed. Participants completed a series of training, gait, and balance tasks in four weeks. The authors measured baseline trunk sway initially. Then they measured trunk sway with vibrotactile feedback using the SwayStar system. There was a substantial reduction in trunk sway (\textit{p} $<$.02) than baseline after one and two weeks of training with vibrotactile feedback. The authors measured a carry-over effect in the fourth week with no training in the third week. They also found a significant carry-over improvement (\textit{p} $<$.02). However, standing with eyes closed on a foam pad provided the best result with 59\% decreased pitch sway (\textit{p} $<$.002) in their study.

Ballardini et al. \cite{ballardini2020vibrotactile} recruited 24 participants (Male: 11, Female: 13) to investigate the effect of vibrotactile feedback on standing balance. They created a device that uses two vibration motors positioned on the anterior and posterior parts of the body to deliver vibrotactile feedback. An accelerometric measurement encoding was synchronized with the vibration that combines the location and acceleration of the body center of mass in the anterior-posterior direction. The purpose was to compare two alternative encoding methods: 1) vibration always on and 2) vibration with a dead zone (i.e., silence when the signal was below the given threshold). Finally, they tested if the informational quality of the feedback influenced these effects by using vibrations unrelated to the real postural oscillations (sham feedback). Nine participants were tested with vibration always on, sham feedback, and fifteen with dead zone feedback. The findings revealed that synchronized vibrotactile feedback reduces sway significantly in the anterior-posterior and medial-lateral directions. The two encoding approaches exhibited no significant difference in lowering postural sway. The presence of sham vibration increased postural sway, emphasizing the relevance of the encoded information. 

While these studies investigated balance in the real world, we applied vibrotactile feedback in VR to find how it affects balance in VR.

\section{Methods}
Our research studied the influence of several vibrotactile feedback (spatial, static, rhythmic, and CoP) on balance in VR settings. On the basis of the past research on assistive technology and VR balance, we explored the following hypotheses:

H1: Each VR-based vibrotactile feedback (spatial, static, rhythmic, and CoP) will considerably enhance balance compared to the no vibration in VR condition.

H2: Spatial vibrotactile feedback will improve balance more effectively than other vibrotactile feedback strategies.

H3: In the absence of vibrotactile feedback in VR, balance will diminish compared to the baseline (non-VR) condition.

\subsection{Participants, Selection Criteria, and Screening Procedure}
To explore the influence of vibrotactile feedback on VR balance, we recruited 39 participants (Male:18, Female:21) from the local area. Six males and twelve females aged 18-75 had BI because of MS. The other twenty-one participants (Male:11, Female:10) did not have BI and did not have MS, arthritis, vestibular dysfunction, or any other medical difficulties; however, they were comparable in age, weight, and height to those with BI. In the BI group, 38.89\% participants identified as White, 44.44\% as Hispanic, 5.56\% as African American, 5.6\% as American Indian, and 5.6\% identified as Asian. 18.18\% of those without BI identified as White, 22.73\% were Hispanic, 50\% were African American, 4.55\% were American Indian, and 4.55\% were Asian. Table 1 displays the mean (SD) values for age, height, weight, and gender for both groups (with and without BI). In our research, more females than men participated. This is due to the fact that we recruited from the MS community, which is statistically more frequent in females \cite{WinNT}. Every participant could walk without support. The participants were recruited from local MS support organizations, rehabilitation centers, and religious communities. The authors' verbal recruitment, email lists, websites, and flyers were the principal methods of recruitment.

\paragraph{Screening Procedure:}

First, we conducted phone interviews with all participants to determine their eligibility for this research. For instance, we first asked them basic questions, such as the year and date (to measure their cognitive abilities) and demographic information. We did not choose anybody who could not comprehend the questions or lacked English proficiency. Then, if known, we inquired about the reasons for their balance issues. We also confirmed that individuals were demographically comparable across populations (i.e., age, height, and weight). Participants who needed medicine to enhance their balance or who could not stand without help were eliminated from the research.

\begin{table}[]
\caption{Descriptive statistics for participants}
    \label{tab:my_label}
\setlength{\tabcolsep}{1pt}
\begin{tabular}{|c|cc|cc|cc|cc|}
\hline
{\textbf{\begin{tabular}[c]{@{}c@{}}Participant \\ Group\end{tabular}}} & \multicolumn{2}{c|}{\textbf{Participants}} & \multicolumn{2}{c|}{\textbf{Age (years)}} & \multicolumn{2}{c|}{\textbf{Height (cm)}} & \multicolumn{2}{c|}{\textbf{Weight (kg)}} \\ \cline{2-9} 
                                                                                       & \multicolumn{1}{c|}{Male}     & Female     & \multicolumn{1}{c|}{Mean}      & SD       & \multicolumn{1}{c|}{Mean}      & SD       & \multicolumn{1}{c|}{Mean}      & SD       \\ \hline
\textbf{BI}                                                                            & \multicolumn{1}{c|}{6}        & 12         & \multicolumn{1}{c|}{39.89}      & 10.18     & \multicolumn{1}{c|}{165.60}    & 10.26    & \multicolumn{1}{c|}{83.86}     & 28.27    \\ \hline
\textbf{Without BI}                                                                    & \multicolumn{1}{c|}{11}        & 10         & \multicolumn{1}{c|}{47.29}      & 12.09     & \multicolumn{1}{c|}{166.12}    & 10.05     & \multicolumn{1}{c|}{88.26}     & 16.05    \\ \hline
\end{tabular}
\end{table}

\subsection{System Description}

\paragraph {Vibrotactile Equipment:}
We used the following vibrotactile equipment from bHaptics (https://www.bhaptics.com):

Vest: Participants wore a wireless vest that included 40 individually controllable ERM (Eccentric Rotating Mass) vibrotactile motors. Twenty of them were on the front side, and the other 20 were on the backside of the vest. The vest was adjustable with shoulder snap buttons. The weight of the vest was 3.7 lbs. 

Arm Sleeves: Participants wore arm sleeves with adjustable straps on both forearms, which were placed in between  the wrist and elbow. Each arm piece had six ERM vibrotactile motors. The weight of each arm piece was 0.66 lbs.

Forehead: We attached it to the HMD to cover around the forehead with six ERM vibrotactile motors. It had a weight of 0.18 lbs.

Fig. 1 shows the positioning of the vibrotactile motors.

\begin{figure}[h!]
\centering
\includegraphics[width=8.25cm,height=3cm]{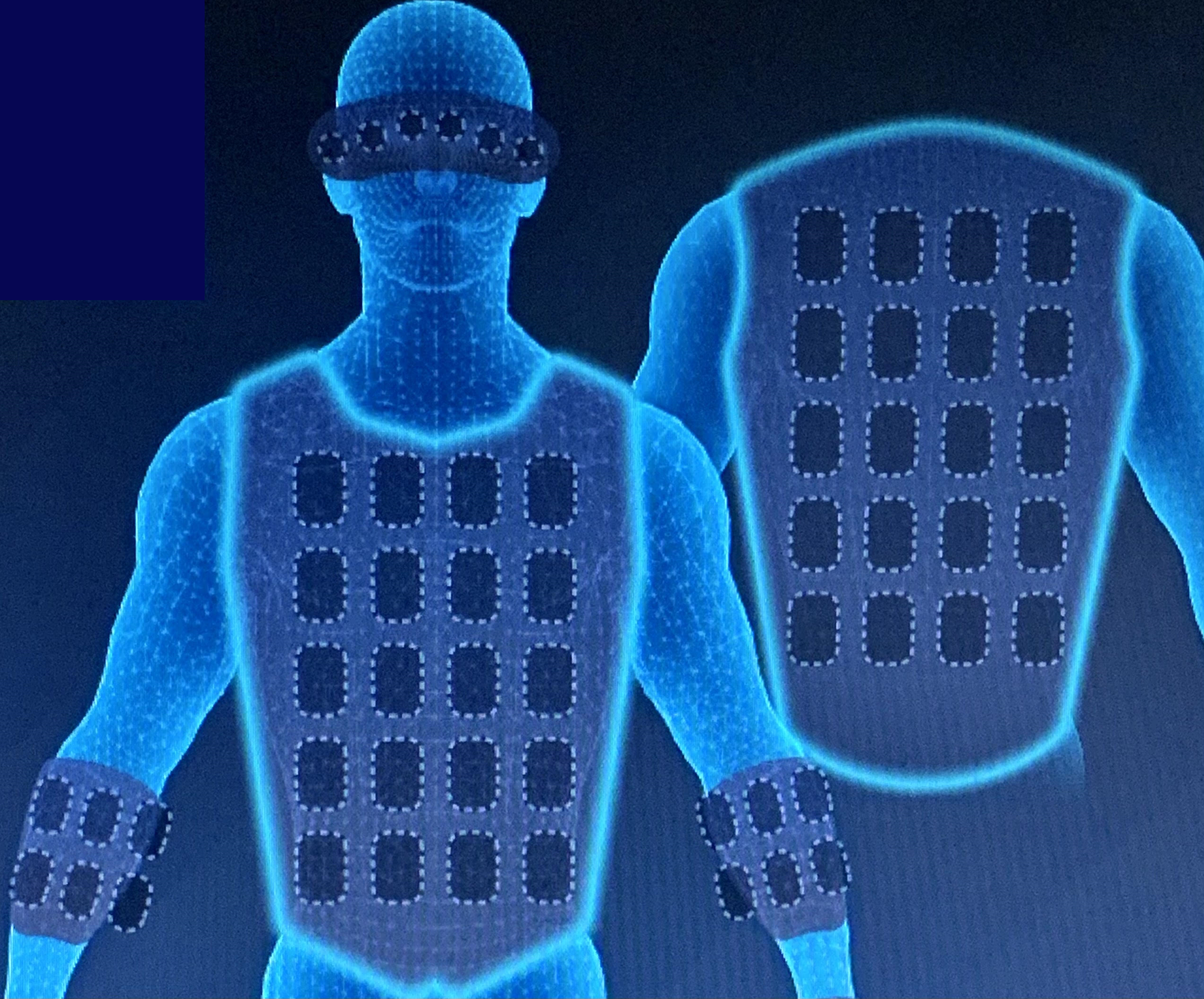}
\caption{ Vibrotactile motors positions.}
\end{figure}

Audio-to-Vibrotactile Software: This software converts audio inputs to corresponding vibrotactile outputs. It also allows for the control of vibration intensity, which we adjusted to the user comfort level in our study.

\paragraph{Balance Measurement:} 
In each condition, the participants’ center of pressure path was measured using the BTrackS Balance Plate (https://balancetrackingsystems.com). The balance plate's sampling frequency was 25 Hz.

\paragraph{Safety Equipment:} 
Participants used a harness to protect them from falling. A partial weight-bearing suspension system was fitted to the harness. Kaye Products Inc. supplied both the harness and the suspension system.

\paragraph{Computers, VR Equipment, and Software:} 
The VEs were developed using Unity3D. We used the HTC Vive Pro Eye, which had a resolution of 2160 x 1200 pixels, a refresh rate of 90 Hz, and a field of view of 110 degrees. We used a Vive tracker affixed to the back of the bHaptics vest to track its position. Vive controllers were used to tracking hand position and orientation while reaching and grasping objects. To render the VE and record the data, we employed a computer with an Intel Core i7 processor (4.20 GHz), 32 GB DDR3 RAM, an NVIDIA GeForce RTX 2080 graphics card, and Windows 10 operating system. We collected the BTrackS Balance Plate data using the NI LabView software (v. 2020) and streamed it to Unity3D via sockets.

\paragraph{Physical Environment:} 
Our lab environment was temperature-controlled and had enough open space ($>$ 600 sq ft.). For the duration of the study, only the participant and the experimenters were permitted in the lab.

\subsection{Study Conditions}
We examined four categories of VR-based vibrotactile feedback approaches and a condition with no vibrotactile feedback to determine how vibrotactile feedback impacts balance in VR. The audio-to-vibrotactile component of the bHaptics program translated the audio into equivalent vibrotactile feedback for each vibrotactile feedback condition. We utilized white noise instead of music or user-selected tones to generate audio since white noise has been demonstrated to alter the signal-to-noise ratio and increase performance owing to the stochastic resonance phenomena \cite{helps2014different}.

\subsubsection{Spatial Vibrotactile Feedback} 
To provide spatial vibrotactile feedback, first, we utilized Google resonance audio SDK in Unity for audio spatialization since the plugin employs head-related transfer functions (HRTFs), and hence this simulates 3D sound more accurately than Unity's default \cite{chong2020audio, pinkl2020spatialized}. The spatial audio in our study was simulated audio (rather than recorded ambisonic audio). The simulated spatialized audio was then sent to the audio-to-vibrotactile software, which generated spatial vibrotactile feedback. The forehead bHaptics device vibrated at varying levels as the user tilted their head. The vest's vibration was modified depending on its location as detected by the Vive tracker. The vibration of the arm sleeves was altered based on the location of the Vive controllers. The X, Y, and Z coordinates of the 3D audio source and the participant in the VE were 0,1,0 and 0,0,0, respectively, which means the audio source was placed just in front of the participant.

\subsubsection{Static Vibrotactile Feedback}  We provided white noise to the bHaptics audio-to-vibrotactile component to generate static vibrotactile feedback. The location of the user had no effect on the feedback. This strategy has also been documented in non-VR research to enhance the balance of adults\cite{ross2016auditory}. 

\subsubsection{Pitch and stereo pan feedback on Center of Pressure (CoP)} 
Similar to static vibrotactile feedback, we transmitted white noise to bHaptics audio-to-vibrotactile software. However, the pitch and stereo pan altered depending on the center of pressure path received from the balance board. We mapped the pitch to the center of pressure from the x coordinate of the balance plate, and stereo pan to the center of pressure from the y coordinate in Unity3D
\cite{hasegawa2017learning}. As a result, when the participant moved from his center position on the balance board to any other side (e.g., left, right, front, or back), the participants felt greater vibration as we designed the vibration intensity to increase with the increase of CoP and vice versa.

\subsubsection{Rhythmic Vibrotactle Feedback} 
At every 1-second interval, we delivered a white noise beat to the bHaptics audio-to-vibrotactile program to produce the rhythmic vibrotactile feedback. The duration of the rhythmic audio clip was also 1-second. Previous research indicated that hearing a constant rhythm may enhance balance and walking in both persons with neurological disorders and older adults in non-VR surroundings \cite{ghai2018effect}.

\subsubsection{No Vibrotactile Feedback} 
This was utilized to evaluate the balance of participants in VR with no vibrotactile input. To maintain consistency with earlier circumstances, participants continued to wear the HMD, bHaptics suit, arm sleeves, and forehead part, but no vibrotactile feedback was provided.

\subsection{Study Procedure}
Figure 2 shows the whole study procedure.

\begin{figure} [h!]

\includegraphics[width=7.4cm,height=8.6cm]{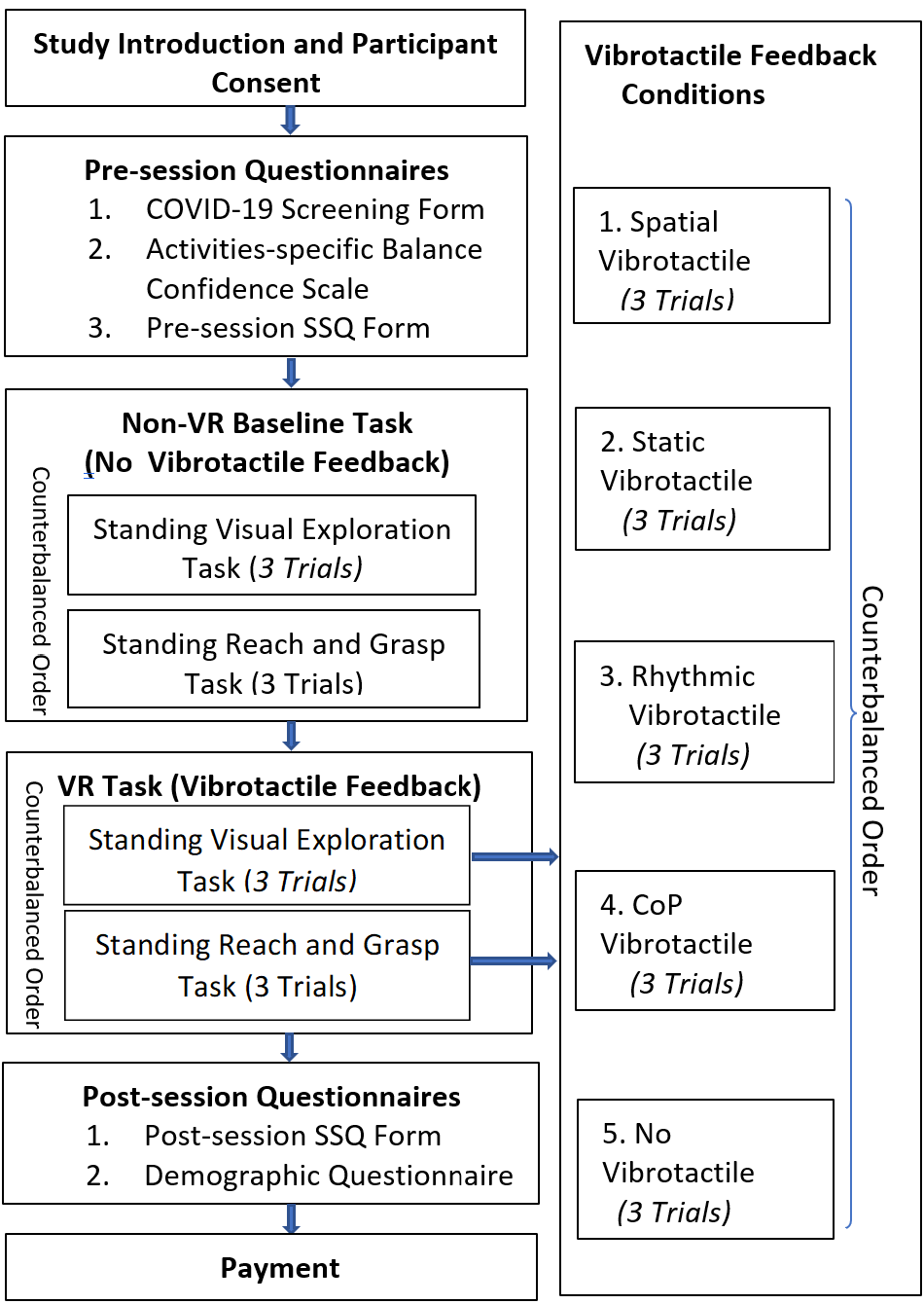}
\caption{Study procedure.}
\end{figure}

The research was authorized by the Institutional Review Board (IRB). Before each user study, we sanitized all equipment (including the HMD, controllers, balance board, objects, harness, and suspension system). The participants completed a COVID-19 screening questionnaire and had their body temperature recorded at the start of the study. The participant then read and signed an informed consent form. We utilized the participants' responses to questions on handedness to find out their dominant and non-dominant hands \cite{coren1993lateral}. Then, we explained the whole study procedure. Participants were then strapped to the harness and suspension system. Throughout the duration of the trial, the participants were supported by the harness, stood on the balance board, and wore no footwear. 

\subsubsection{Pre-Session Questionnaires} The participants completed an Activities-specific Balance Confidence (ABC) \cite{schepens2010short} and a Simulator Sickness Questionnaire (SSQ) \cite{kennedy1993simulator} at the beginning of the study.

\subsubsection{Tasks} 
Participants completed a visual exploration task and a reach and grasp task while standing. Tasks were performed in both VR settings and a non-VR environment. VEs were simulations of the actual settings. To minimize the likelihood of confounding factors, such as the learning effect, we ensured that both tasks were done in a counterbalanced sequence.

\paragraph{Standing Visual Exploration Task:} At the beginning of each condition and trial, a prerecorded instruction led participants to gaze at markers placed in various positions across the room. The markers had directional designations, and the instruction instructed them to observe a certain marker each time. There was a two-second wait between each direction in the instruction. The time was three minutes in total. In the laboratory, we set several markers — 'Left,' 'Right,' 'Top,' 'Bottom,' and 'Front' — in their respective directions. We monitored the head movements and generated pictures of the participants to confirm they were following the instructions. We wanted all participants to view the laboratory in a standardized manner to ensure consistency. During their exploration of the VE, the participants stood straight on the balance board and were not permitted to move their bodies except for their heads. We gathered data in real-time from the BTrackS balance plate. Figure 3 shows the comparison between baseline and VR tasks for the standing visual exploration task. The actual environment and associated VE have been shown in Figure 4. To design the standing visual exploration task, we reproduced the task provided in \cite{ferdous2018investigating}. We selected to execute this motor activity in a laboratory VE because we wanted to compare their balance in the VE lab to their balance in the actual lab.

\begin{figure}[h!]
\includegraphics[width=8.5cm,height=3.7cm]{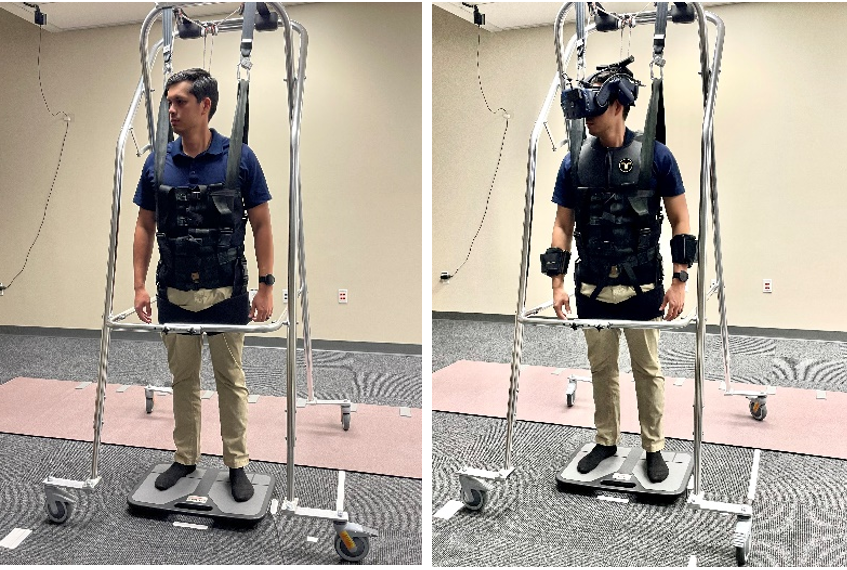}
\caption{Participant performing baseline (left) and VR task (right) for standing visual exploration task.}
\end{figure}

\begin{figure}[h!]
\includegraphics[width=4.25cm,height=3.7cm]{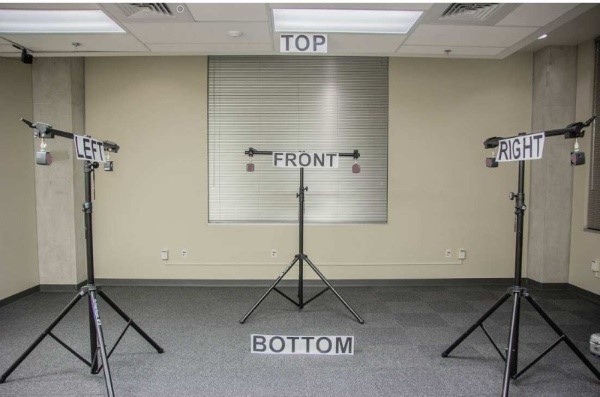}
\includegraphics[width=4.25cm,height=3.7cm]{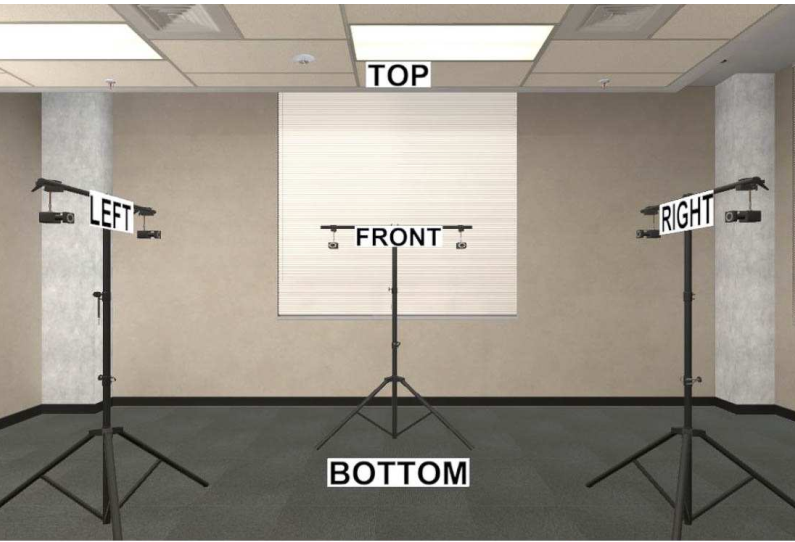}
\caption{Standing visual exploration task: real environment (left) and virtual environment (right).}
\end{figure}

Standing Reach and Grasp Task: Participants grasped actual items that were within their reach. We positioned four items (cubes - 5.08 cm) on the table at the designated locations. The distance between each pair of items was 24 cm. The balance board was positioned on the ground in the center of the table. The distance between the table and balance board was 12 cm. Participants were barefoot and placed each foot on the balance board's specified locations. The participants rested their non-dominant hands on their upper thighs and reached and grasped the items with their dominant hands. Participants were permitted to reach the items by leaning forward to their utmost comfortable distance without taking their heels off the balancing board and standing upright. The participants were directed to grasp the four items in a random sequence, lift them to chest level, and then return them to their original location. To accomplish this objective, we followed the task in \cite{cordova2014older}. Fig. 5 depicts the workspace and a comparison of the actual world and virtual environment for this activity. We selected this motor task because reaching is an essential element of daily life, it has been used to test balance, and it is frequent in VR \cite{bolton2019motor, huang2015effects, tan2012anticipatory}.

\begin{figure}[h!]
\centering
\includegraphics[width=8.25cm,height=3.5cm]{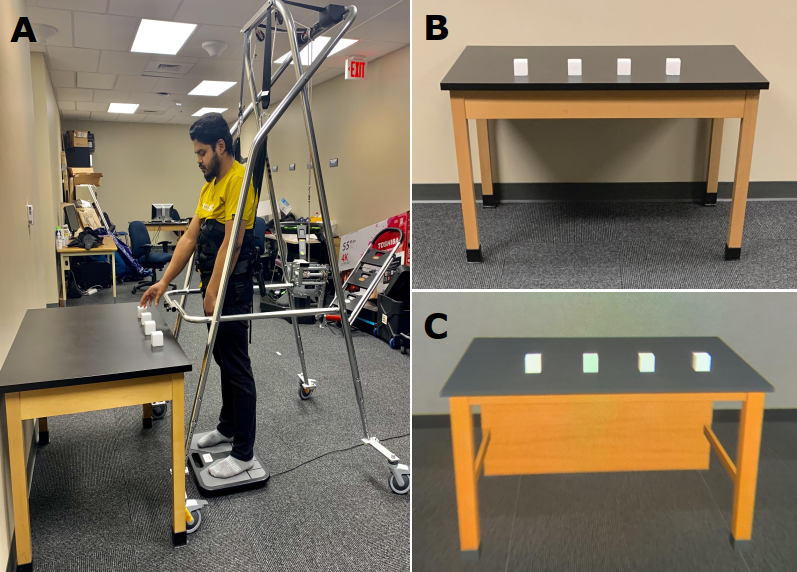}
\caption{Standing reach and grasp task: (A) Workspace (B) Real environment (C) Virtual environment}
\end{figure}

\subsubsection{Baseline Measurements without VR} The participants stood on a BTrackS Balance Plate while a harness prevented them from falling. Then, we assessed their balance throughout three trials of standing visual exploration and standing reach and grasp tasks. Each trial lasted for three minutes.

\subsubsection{VR Tasks}  All of these are repetitions of the aforementioned baseline activities, with the exception that they were conducted in VR, resulting in small changes. For both virtual activities, participants utilized the HTC Vive HMD to view the virtual environment. We did the following tasks under four distinct vibrotactile feedback and a VR condition with no vibration and three times in each session while we counterbalanced the vibrotactile conditions and tasks.

\paragraph{Standing Visual Exploration Task in VR:}  To explore the VE, participants followed the same recorded instructions used for real-world balance assessments. The same size virtual markers were set in the same position as the baseline assessment. The measurement procedure was the same as in the standing visual exploration task used in the baseline.

\paragraph{Standing Reach and Grasp Task in VR:}  Using the controllers, participants reached for and grasped virtual items within their reach with their dominant hand. When participants touched the virtual items with the controller, the color of the object changed from blue to red. The participants then pressed the trigger on the controller to grasp the items, lifted them to chest level, and released the trigger when returning the item to its original location. The virtual environment and measures were identical to the baseline task.


\subsection{Post-Session Questionnaires}
At the end of the study, participants completed an SSQ and a demographic questionnaire.

Participants required approximately two hours to complete the study. At the end of the study, each participant was compensated 30 USD per hour and given money for parking.

\section{Metrics}
\subsection{Center of Pressure (CoP) Velocity}
In our research, CoP velocity was the key parameter of balance measurement. We selected CoP velocity because it is generally recognized as a good measurement for evaluating balance \cite{li2016reliability}. Using the following formula by Young et al. \cite{young2011assessing}, we computed CoP from the four pressure sensors on the BTrackS Balance Plate.

\begin{equation}
CoP (X,Y) = \frac{\sum_{i=1}^{4} Weight_{i} * (x_{i},y_{i})} {\sum_{i=1}^{4} Weight_{i}} 
\end{equation}

Where $(x_{i},y_{i})$ = coordinates of the pressure sensor \textit{i}, $Weight_{i}$ = weight or pressure data on the $i$th sensor, and $CoP(X,Y)$ = coordinates of the $CoP$.

Then, we computed the CoP path for all samples using the following formula.

\begin{equation}
CoP\;Path = \sum_{i=1}^{n-1}\sqrt{(CoP_{i+1}X - CoP_{i}X)^2 + (CoP_{i+1}Y - CoP_{i}Y)^2}
\end{equation}

Here, $CoP_{i}X$ = $X$ coordinate of CoP at $i$th second, and $CoP_{i}y$ = $Y$ coordinate of CoP at $i$th second.

Finally, we calculated CoP velocity by dividing the CoP path for all samples by the total data recording time for all samples (T).

\begin{equation}
    CoP\;Velocity = \frac{CoP\;Path}{T}
\end{equation}

\subsection{Activities-specific Balance Confidence (ABC) Scale}
ABC is a 16-item questionnaire in which each question inquires about the participant's confidence in doing a particular daily living activity \cite{powell1995activities}. The ABC score is computed by adding the percentages from each question (1-16), with a maximum score of 1600 possible points. The ABC percent is calculated by dividing the total by 16.

\subsection{Simulator Sickness Questionnaire}
The Simulator Sickness Questionnaire (SSQ) is a 16-item questionnaire in which each question inquires about the physiological discomfort of the participant \cite{kennedy1993simulator}. This test is necessary for identifying individuals prone to severe cybersickness and examining the link with postural instability.

\section{Statistical Analysis}
 We used the Shapiro-Wilk test to examine the normality of the data. The data for participants with and without BI were normally distributed for both tasks; $p$ =.321, $w$ = 0.89. Then, we conducted a 2$\times$6 mixed-model ANOVA with two between-subject factors (participants with BI and participants without BI) and six within-subject factors (six study conditions: baseline, spatial, static, rhythmic, CoP, and no audio) to identify any significant difference in CoP velocities. When there was a significant difference, post-hoc two-tailed t-tests were performed for comparisons of within and between groups. For cybersickness analysis, we also used two-tailed t-tests comparing pre-session and post-session SSQ scores for each participant group. Additionally, we conducted two-tailed t-tests comparing the ABC scores of both participant groups to assess the difference in physical ability. Bonferroni correction was used for all tests involving multiple comparisons.
 
\section{Results}
We compared CoP velocities between study conditions and obtained the following results.
\subsection{Within Group Comparisons on CoP Velocity}
ANOVA tests revealed substantial difference for individuals with BI, \textit{F}(1,123) = 19.6, \textit{p} $<$ .001; and effect size, $\eta^{2}$ = 0.08 for standing visual exploration task and \textit{F}(1,123) = 51.3, \textit{p} $<$ .001; and effect size, $\eta^{2}$ = 0.07 for standing reach and grasp task. We also found substantial difference for individuals without BI, \textit{F}(1,123) = 18.02, \textit{p} $<$ .001; and effect size, $\eta^{2}$ = 0.06 for standing visual exploration task and \textit{F}(1,123) = 41.72, \textit{p} $<$ .001; and effect size, $\eta^{2}$ = 0.05 for standing reach and grasp task. Next, for each group separately, we conducted the following pair-wise comparisons applying two-tailed t-tests to identify differences between specific study conditions.

\subsubsection{Baseline vs. No Vibrotactile Feedback}
\paragraph{Standing Visual Exploration Task:}
Experiment results showed a significant difference between no vibrotactile (Mean, \textit{M} = 4.44, Standard Deviation, \textit{SD} = 1.64) and baseline (\textit{M}= 3.32, \textit{SD} = 0.95) condition; \textit{t}(17) = 3.57, \textit{p} = .002, \textit{r} = 0.6 for participants with BI. 
Similarly, we observed a significant difference between no vibrotactile (\textit{M} = 4.33, \textit{SD} = 1.32) and baseline (\textit{M}= 3.66, \textit{SD} = 1.33) condition; \textit{t}(20) = 3.28, \textit{p} = .004, \textit{r} = 0.76 for participants without BI.

\paragraph{Standing Reach and Grasp Task:}
We obtained a significant difference between no vibrotactile (\textit{M} = 6.53, \textit{SD} = 1.37) and baseline (\textit{M}= 5.5, \textit{SD} = 1.42) condition; \textit{t}(17) = 3.5, \textit{p} = .003, \textit{r} = 0.6 for participants with BI.
We also found a significant difference between no vibrotactile (\textit{M} = 5.63, \textit{SD} = 1.17) and baseline (\textit{M}= 4.98, \textit{SD} = 1.07) condition; \textit{t}(20) = 3.43, \textit{p} = .003, \textit{r} = 0.71 for participants without BI.
Thus, we observed a significant increase of CoP velocity in no vibrotactile in VR than the baseline condition in this case. 


\begin{figure}[ht!]
    \centering
  \includegraphics[width=0.5\textwidth,height=4.5cm]{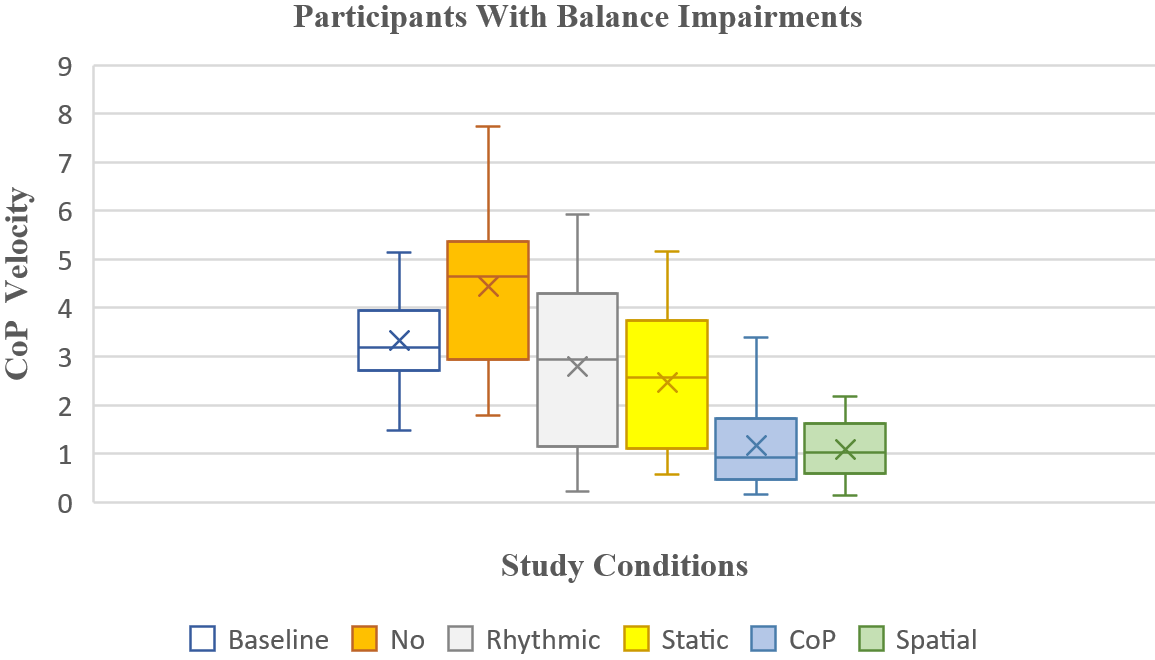}
\break
   \includegraphics[width=0.5\textwidth,height=4.5cm]{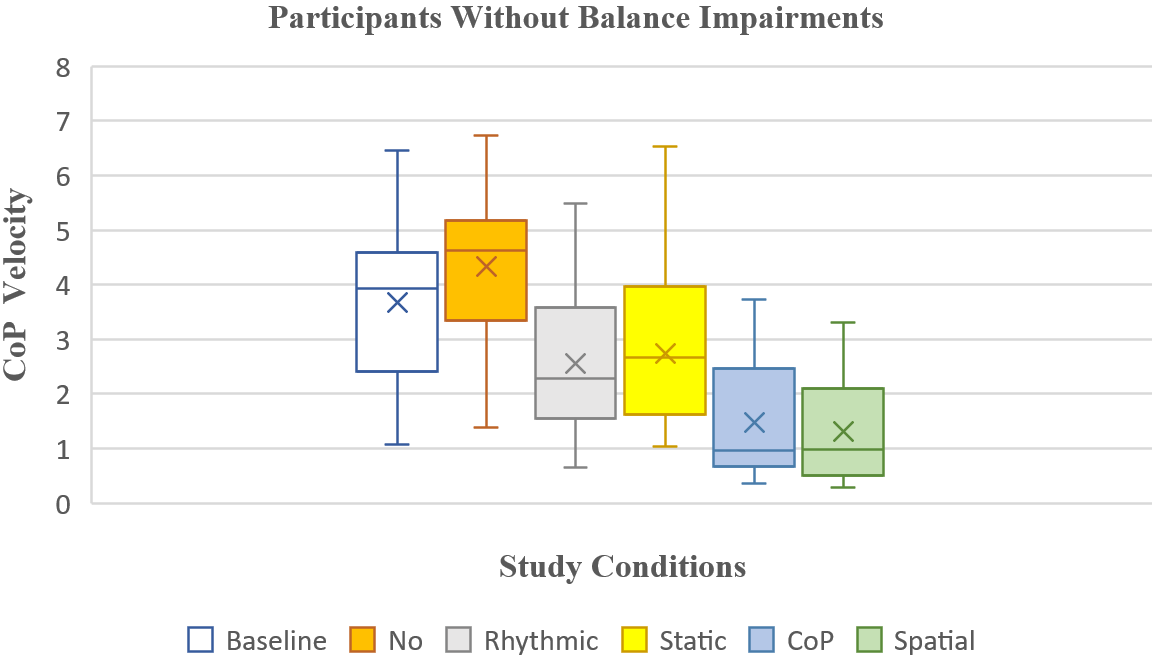}
  \caption{CoP velocity comparison between study conditions for standing visual exploration task.}
\end{figure}

\subsubsection{No Vibrotactile vs. Spatial Vibrotactile Feedback}
\paragraph{Standing Visual Exploration Task:}
For participants with BI, CoP velocity was significantly reduced in the spatial condition (\textit{M} = 1.08, \textit{SD} = 0.64 ) compared to no vibrotactile condition; \textit{t}(17) = 10.96, \textit{p} $<$ .001, \textit{r} = 0.68.
For individuals without BI, CoP velocity was substantially lower in spatial condition (\textit{M} = 1.32, \textit{SD} = 1.33) than in the no vibrotactile condition; \textit{t}(20) = 10.01, \textit{p} $<$ .001, \textit{r} = 0.30. 

\paragraph{Standing Reach and Grasp Task:}
For individuals with BI, the resulting CoP velocity was substantially slower in spatial (\textit{M} = 2.01, \textit{SD} = 0.96 ) than in the absence of vibrotactile condition; \textit{t}(17) = 13.66, \textit{p} $<$ .001, \textit{r} = 0.32.
CoP velocity was substantially less in spatial (\textit{M} = 2.18, \textit{SD} = 0.88) than the absence of vibrotactile condition for individuals without BI; \textit{t}(20) = 10.92, \textit{p} $<$ .001, \textit{r} = 0.03.
Thus, spatial vibrotactile outperformed no vibrotactile in VR for both standing visual exploration and standing reach and grasp tasks.

\subsubsection{ No Vibrotactile vs. CoP Vibrotactile Feedback}
\paragraph{Standing Visual Exploration Task:}
CoP condition (\textit{M} = 1.17, \textit{SD} = 0.91) indicated substantial improvement of balance relative to the absence of vibrotactile condition; \textit{t}(17) = 9.84, \textit{p} $<$ .001, \textit{r} = 0.52 for individuals with BI. 
CoP velocity was also substantially reduced in CoP vibrotactile (\textit{M} = 1.48, \textit{SD} = 0.97 ) contrast to the absence of vibrotactile feedback in VR; \textit{t}(20) = 10.38, \textit{p} $<$ .001, \textit{r} = 0.44 for persons without BI.

\paragraph{Standing Reach and Grasp Task:}
We observed that CoP velocity was significantly reduced in CoP vibrotactile (\textit{M} = 2.11, \textit{SD} = 1.13) compared to the absence of vibrotactile feedback in VR; \textit{t}(17) = 9.91, \textit{p} $<$ .001, \textit{r} = 0.13 for individuals with BI.
CoP velocity was also substantially reduced in CoP vibrotactile (\textit{M} = 2.4, \textit{SD} = 0.93 ) relative to the absence of vibrotactile feedback in VR; \textit{t}(20) = 9.67, \textit{p} $<$ .001, \textit{r} = 0.04 for persons without BI.
Therefore, the CoP vibrotactile condition performed significantly better than the no vibrotactile feedback in VR for both tasks.

\subsubsection{No Vibrotactile vs. Static Vibrotactile Feedback}
\paragraph{Standing Visual Exploration Task:}
The obtained CoP velocity was substantially reduced in static (\textit{M} = 2.47, \textit{SD} = 1.38 ) compared to the absence of vibrotactile condition; \textit{t}(17) = 7.07, \textit{p} $<$ .001, \textit{r} = 0.71 for individuals with BI. 
Experimental results also revealed that CoP velocity was considerably less in static vibrotactile (\textit{M} = 2.75, \textit{SD} = 1.36 ) relative to the absence of vibrotactile feedback for persons without BI; \textit{t}(20) = 7.29, \textit{p} $<$ .001, \textit{r} = 0.72.

\paragraph{Standing Reach and Grasp Task:}
We found that CoP velocity was considerably reduced in static (\textit{M} = 4.14, \textit{SD} = 1.3) contrast to the absence of vibrotactile feedback in VR; \textit{t}(17) = 6.31, \textit{p} $<$ .001, \textit{r} = 0.28 for individuals with BI.
CoP velocity was also significantly lower in static vibrotactile (\textit{M} = 3.71, \textit{SD} = 1.29) compared to the absence of vibrotactile condition; \textit{t}(20) = 5.77, \textit{p} $<$ .001, \textit{r} = 0.24 for persons without BI.
Hence, static vibrotactile outperformed the no vibrotactile in VR conditions for both tasks.

\subsubsection{ No Vibrotactile vs. Rhythmic Vibrotactile Feedback}
\paragraph{Standing Visual Exploration Task:}
We noticed that CoP velocity was substantially reduced in rhythmic (\textit{M} = 2.79, \textit{SD} = 1.83) compared to the absence of vibrotactile feedback for individuals with BI; \textit{t}(17) = 10.77, \textit{p} $<$ .001, \textit{r} = 0.91. 
CoP velocity was also considerably reduced in rhythmic vibrotactile (\textit{M} = 2.55, \textit{SD} = 1.33) compared to the absence of vibrotactile feedback for persons without BI; \textit{t}(20) = 6.98, \textit{p} $<$ .001, \textit{r} = 0.61.

\paragraph{Standing Reach and Grasp Task:}
Experimental results showed that CoP velocity was substantially diminished in rhythmic (\textit{M} = 4.04, \textit{SD} = 1.6) compared to the absence of vibrotactile feedback in VR for individuals with BI; \textit{t}(17) = 6.7, \textit{p} $<$ .001, \textit{r} = 0.44.
CoP velocity was also considerably less in rhythmic vibrotactile (\textit{M} = 3.58, \textit{SD} = 1.42 ) relative to the absence of vibrotactile feedback for persons without BI; \textit{t}(20) = 5.59, \textit{p} $<$ .001, \textit{r} = 0.16. 
Therefore, rhythmic vibrotactile performed better than the no vibrotactile feedback in VR.

\subsubsection{ Rhythmic Vibrotactile vs. Spatial Vibrotactile Feedback}
\paragraph{Standing Visual Exploration Task:}
CoP velocity was substantially diminished in spatial compared to rhythmic vibrotactile for both individuals with BI (\textit{t}(17) = 4.74, \textit{p} $<$ .001, \textit{r} = 0.6) and for individuals without BI (\textit{t}(20) = 4.09, \textit{p} $<$ .001, \textit{r} = 0.30). 

\paragraph{Standing Reach and Grasp Task:}
Experimental results also revealed that CoP velocity was considerably decreased in spatial relative to rhythmic vibrotactile for individuals with BI (\textit{t}(17) = 5.75, \textit{p} $<$ .001, \textit{r} = 0.41) and for individuals without BI (\textit{t}(20) = 5.52, \textit{p} $<$ .001, \textit{r} = 0.58).
Therefore, spatial vibrotactile performed better than rhythmic vibrotactile to improve balance in our study.

\begin{figure}[ht!]
    \centering
  \includegraphics[width=0.5\textwidth,height=4.5cm]{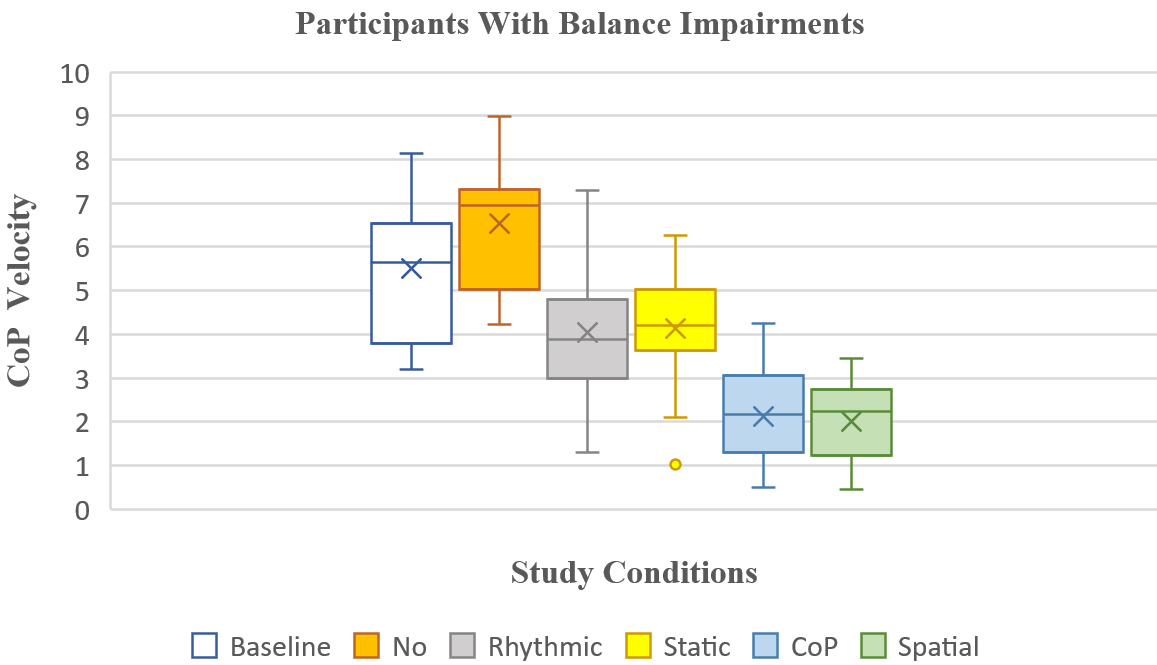}
\break
  \includegraphics[width=0.5\textwidth,height=4.5cm]{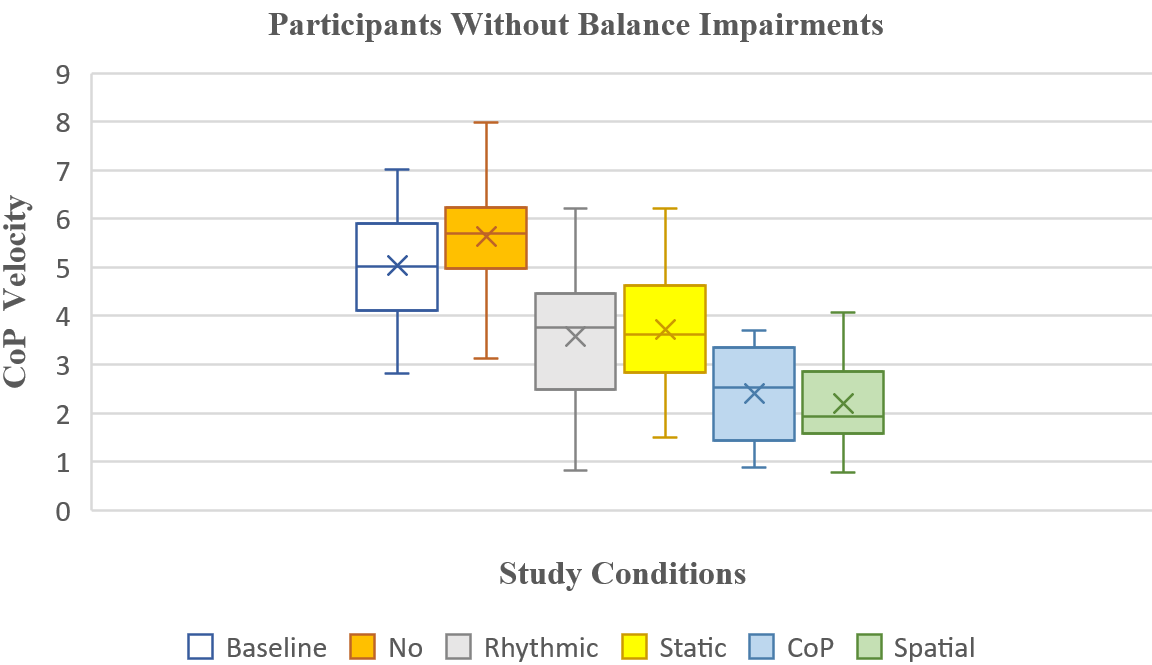}
  \caption{CoP velocity comparison between study conditions for standing reach and grasp task.}
\end{figure}

\subsubsection{ Rhythmic Vibrotactile vs. CoP Vibrotactile Feedback}
\paragraph{Standing Visual Exploration Task:}
CoP velocity was substantially reduced in CoP vibrotactile than rhythmic vibrotactile for individuals with BI (\textit{t}(17) = 4.55, \textit{p} $<$ .001, \textit{r} = 0.17) and for individuals without BI (\textit{t}(20) = 5.12, \textit{p} $<$ .001, \textit{r} = 0.69).

\paragraph{Standing Reach and Grasp Task:}
We also noticed significantly lower CoP velocity in CoP vibrotactile than rhythmic for individuals with BI (\textit{t}(17) = 4.34, \textit{p} $<$ .001, \textit{r} = 0.49) and for individuals without BI (\textit{t}(20) = 5.14, \textit{p} $<$ .001, \textit{r} = 0.67). These results suggested that CoP vibrotactile can be preferred over rhythmic vibrotactile to reduce imbalance issue in VR.

\subsubsection{ Rhythmic Vibrotactile vs. Static Vibrotactile Feedback}
\paragraph{Standing Visual Exploration Task:}
The obtained CoP velocity indicated no significant difference between rhythmic and static vibrotactile  for individuals with BI (\textit{t}(17) = 1.1, \textit{p} = .288, \textit{r} = 0.74) and for individuals without BI (\textit{t}(20) = 0.99, \textit{p} = .332, \textit{r} = 0.77).

\paragraph{Standing Reach and Grasp Task:}
For individuals with and without BI, there was no significant difference between rhythmic and static vibrotactile scenarios.
For BI, \textit{t}(17) = 0.43, \textit{p} = .67, \textit{r} = 0.77 and for without BI, \textit{t}(20) = 0.96, \textit{p} = .344, \textit{r} = 0.89.
Therefore, there was no clear indication of which vibrotactile feedback could be chosen between rhythmic and static to improve balance.

\subsubsection{ Static Vibrotactile vs. Spatial Vibrotactile Feedback}
\paragraph{Standing Visual Exploration Task:}
We found that CoP velocity was substantially less in spatial compared to static vibrotactile scenario for both individuals with and without BI. For BI, \textit{t}(17) = 6.01, \textit{p} $<$ .001, \textit{r} = 0.76. For without BI, \textit{t}(20) = 5.69, \textit{p} $<$ .001, \textit{r} = 0.55.

\paragraph{Standing Reach and Grasp Task:}
For both individuals with and without BI, experimental results showed that CoP velocity was considerably slower in spatial relative to static vibrotactile scenario. For individuals with BI, \textit{t}(20) = 8.31, \textit{p} $<$ .001, \textit{r} = 0.41. For individuals without BI, \textit{t}(20) = 6.48, \textit{p} $<$ .001, \textit{r} = 0.56.
Thus, spatial vibrotactile performed better than static in VR to improve balance.

\subsubsection{ Static Vibrotactile vs. CoP Vibrotactile Feedback}
\paragraph{Standing Visual Exploration Task:}
For both individuals with and without BI, experimental results revealed that CoP velocity was substantially decreased in CoP vibrotactile than static vibrotactile feedback. For individuals with BI, \textit{t}(17) = 5.65, \textit{p} $<$ .001, \textit{r} = 0.71. For individuals without BI, \textit{t}(20) = 6.86, \textit{p} $<$ .001, \textit{r} = 0.79.

\paragraph{Standing Reach and Grasp Task:}
We also observed sunstantial decrease in CoP velocity in CoP vibrotactile condition compared to static condition for both group of participants. For individuals with BI, \textit{t}(17) = 7.26, \textit{p} $<$ .001, \textit{r} = 0.43. For individuals without BI,\textit{t}(20) = 5.48, \textit{p} $<$ .001, \textit{r} = 0.55.
Thus, CoP vibrotactile feedback outperformed static conditions for both tasks.

\subsubsection{ CoP Vibrotactile vs. Spatial Vibrotactile Feedback}
\paragraph{Standing Visual Exploration Task:}
No substantial difference was found between spatial and CoP vibrotactile condition for both individuals with BI (\textit{t}(17) = 0.55, \textit{p} = .586, \textit{r} = 0.77) and individuals without BI (\textit{t}(20) = 0.67, \textit{p} = .506, \textit{r} = 0.3).

\paragraph{Standing Reach and Grasp Task:}
Experimental results did not show a considerable difference between spatial and CoP vibrotactile condition for individuals with BI (\textit{t}(17) = 0.31, \textit{p} = .756, \textit{r} = 0.32) and for individuals without BI (\textit{t}(20) = 0.96, \textit{p} = .343, \textit{r} = 0.38).
As a result, it was unclear which vibrotactile feedback could be chosen between spatial and CoP to solve gait disturbance issues in VR environments.

\begin{table}[t]
\caption{Summarized results for pairwise comparisons}
    \label{tab:my_label}
    \setlength{\tabcolsep}{1.5pt}
\begin{tabular}{|@{}l@{}|@{}cc@{}|@{}cc|}
\hline
\multicolumn{1}{|c|}{\textbf{Comparisons }} & \multicolumn{2}{c|}{\textbf{\begin{tabular}[c]{@{}c@{}}Standing Visual \\ Exploration\end{tabular}}} & \multicolumn{2}{c|}{\textbf{\begin{tabular}[c]{@{}c@{}}Standing Reach\\  and Grasp\end{tabular}}} \\ \cline{2-5} 
                                      & \multicolumn{1}{c|}{BI}                    & \begin{tabular}[c]{@{}c@{}}Without BI\end{tabular}   & \multicolumn{1}{c|}{BI}                  & \begin{tabular}[c]{@{}c@{}}Without BI\end{tabular}  \\ \hline
\textbf{Spatial vs. Static}           & \multicolumn{1}{c|}{p \textless .001}      & p \textless .001                                        & \multicolumn{1}{c|}{p \textless .001}    & p \textless .001                                       \\ \hline
\textbf{Spatial vs. Rhythmic}         & \multicolumn{1}{c|}{p \textless .001}      & p \textless .001                                        & \multicolumn{1}{c|}{p \textless .001}    & p \textless .001                                       \\ \hline
\textbf{Spatial vs. CoP}              & \multicolumn{1}{c|}{p \textgreater .05}      & p \textgreater .05                                        & \multicolumn{1}{c|}{p \textgreater .05}    & p \textgreater .05                                       \\ \hline
\textbf{Spatial vs. No vibro}               & \multicolumn{1}{c|}{p \textless .001}      & p \textless .001                                        & \multicolumn{1}{c|}{p \textless .001}    & p \textless .001                                       \\ \hline
\textbf{Static vs. Rhythmic}          & \multicolumn{1}{c|}{p \textgreater .05}    & p \textgreater .05                                      & \multicolumn{1}{c|}{p \textgreater .05}  & p \textgreater .05                                     \\ \hline
\textbf{Static vs. CoP}               & \multicolumn{1}{c|}{p \textless .001}      & p \textless .001                                        & \multicolumn{1}{c|}{p \textless .001}    & p \textless .001                                       \\ \hline
\textbf{Static vs. No vibro}                & \multicolumn{1}{c|}{p \textless .001}      & p \textless .001                                        & \multicolumn{1}{c|}{p \textless .001}    & p \textless .001                                       \\ \hline
\textbf{Rhythmic vs. CoP}             & \multicolumn{1}{c|}{p \textless .001}      & p \textless .001                                        & \multicolumn{1}{c|}{p \textless .001}    & p \textless .001                                       \\ \hline
\textbf{Rhythmic vs. No vibro}              & \multicolumn{1}{c|}{p \textless .001}      & p \textless .001                                        & \multicolumn{1}{c|}{p \textless .001}    & p \textless .001                                       \\ \hline
\textbf{CoP vs. No vibro}                   & \multicolumn{1}{c|}{p \textless .001}      & p \textless .001                                        & \multicolumn{1}{c|}{p \textless .001}    & p \textless .001                                       \\ \hline
\textbf{Baseline vs. No vibro}              & \multicolumn{1}{c|}{p \textless .05}    & p \textless .05                                      & \multicolumn{1}{c|}{p \textless .05}  & p \textless .05                                     \\ \hline
\end{tabular}
\end{table}
Fig. 5 and Fig. 6 represent the experimental results for the standing visual exploration task and standing reach and grasp task, respectively. Table 2 represents the summarized results.

\subsection{Between Group Comparisons}
Results from mixed-model ANOVA and post-hoc two-tailed t-tests indicated that there was a significant difference in CoP velocities for baseline conditions between the two groups (participants with and without BI); \textit{t}(38) = 8.31, \textit{p} $<$ .001, \textit{r} = 0.41.
However, we did not observe any significant difference between other study conditions.

\subsection{ABC Scale}
We administered ABC scale to all participants where 80\% indicates high level, 50-80\% indicates moderate level, and $<$ 50\% indicates poor level of physical functioning. We performed a two-tailed t-test between the ABC score of the participants with BI ($M$ = 63.28, $SD$ = 19.36) and those without BI ($M$ = 93.67, $SD$ = 9.98); \textit{t}(38) = 4.11, \textit{p} $<$ .001. The mean ABC score of the participants with BI was 63.28\%, which suggested the participants with BI had a moderate level of physical functioning. In contrast, the mean ABC score of the participants without BI was 93.67\%, which demonstrated their high level of physical functioning. 

\subsection{Simulator Sickness Questionnaire}
We conducted a two-tailed t-test comparing the pre-session SSQ score and the post-session SSQ score for individuals with and without BI. There was no substantial change between the pre-session and post-session SSQ scores for both participant groups.  We found \textit{t}(17) = 1.39, \textit{p} = .07, \textit{r} = 0.8 for individuals with BI, while \textit{t}(20) = 1.18, \textit{p} = .09, \textit{r} = 0.49 for individuals without BI.

\section{Discussion}
\subsection{Effect of Vibrotactile Feedback on Balance}
Balance increases when CoP velocity decreases \cite{thompson2017balance,ruhe2011center}. Our experimental findings revealed the following effect of vibrotactile feedback on balance.

\subsubsection{No Vibrotactile in VR vs. All VR Based Vibrotactile Feedback:}
CoP velocity was significantly (\textit{p} $<$ .001) slower throughout all vibrotactile feedback compared to the no vibration in VR scenario for participants with and without BI for both tasks (standing visual exploration and standing reach and grasp). Thus, spatial, CoP, rhythmic, and static vibration substantially improved balance for both participants with and without BI. As a result, H1 can be accepted. 


However, the results of this study suggested a divergence from the past research \cite{ferdous2018investigating}. They explored the influence of visual feedback on balance using the standing visual exploration task for both individuals with and without BI in VR. They observed that visual feedback enhanced balance in individuals with BI but found no significant effects of visual conditions on balance in those without BI. Interestingly, based on our experimental findings, vibrotactile feedback improved balance for both people with and without BI. In earlier research \cite{baram2012gait}, auditory input was also shown to be more useful than visual feedback during VR walking.

\subsubsection{ Comparisons of All VR Based Vibrotactile Feedback:}
In all tasks of standing visual exploration and standing reach and grasp, spatial and CoP vibrotactile feedback performed significantly (\textit{p} $<$ .001) better than rhythmic and static conditions for individuals with and without BI which supported our hypothesis H2. There was no statistically significant difference between groups in the spatial and CoP vibrotactile conditions. Also, there was no significant difference between static and rhythmic vibrotactile feedback conditions. This might be due to the fact that both activities tested were stationary and used a simple VE. A study with more participants and more complex VEs might find a significant difference between spatial and CoP or between static and rhythmic conditions.



\subsection{Spatial and CoP Vibrotactile Feedback}
According to the findings, spatial and CoP vibration enhanced balance substantially more than other vibrotactile conditions. When the participants leaned slightly in either direction while standing on the balance board or tilted their heads, the spatial and CoP vibration intensities changed. As these two vibrotactile circumstances worked substantially better than the other vibrotactile conditions, we hypothesized that the vibration level that varies based on the participant's posture gives significantly more effective feedback to participants for adjusting their posture.

\subsection{Between Group Comparisons}
The ABC scores revealed that the physical functioning of participants with BI was significantly lower than that of people without BI. The difference in CoP velocities between the two groups' baseline conditions was also significant. However, there was no statistically significant difference across the remaining VR scenarios. For additional analysis, we subtracted baseline data from all conditions to determine which group had the greatest improvement in balance. The results of a mixed ANOVA and post-hoc two-tailed t-tests across the two groups revealed that participants with BI improved their balance more than those without BI. We hypothesized that since participants with BI had impaired balance function, they would have a greater likelihood of improving their balance in VR than people without BI. This finding was also supported by previous research in which it was discovered that people with BI improved their balance and gait substantially more than those without BI \cite{guo2015mobility}. Because the individuals with BI improved their balance more than the people without BI in VR, this could have been one reason why there was no significant difference between the two groups in VR despite a considerable difference in baseline circumstances. However, more research is needed to confirm this.

\subsection{Effect of Virtual Environment}
We discovered that the CoP velocity was significantly increased (\textit{p} < .05) in the no vibration in VR condition compared to the baseline condition for both standing visual exploration and standing reach and grasp tasks in both participant groups. As balance diminish with the increase of CoP velocity, our hypothesis H3 was supported. Prior studies revealed that postural instability often rises in VR primarily because of sensory conflicts \cite{epure2014effect,soffel2016postural}, resulting in greater CoP velocity in VR than in a baseline condition. 

\subsection{Cybersickness}
We did not observe a significant difference (\textit{p} = .35) between the pre-session and post-session SSQ scores. Participants may have experienced some cybersickness due to the fact that our research consisted of two distinct tasks, six distinct conditions for each task, and three trials for each condition, which took approximately two hours to complete the study. Cybersickness is frequent in VR activities lasting more than 10 minutes \cite{chang2020virtual,kim2021clinical}. However, since there was no illusion of self-motion in our VEs, our setting was tailored to reduce cybersickness. Thus, we expect that cybersickness did not have a significant influence on the CoP velocity findings.

\subsection{Limitations}
The baseline measurement could be expanded. At the outset of the study, participants were informed of the whole research protocol. Then, we conducted several trials with them until they were used to the experimental techniques. Additionally, we conducted three baseline trials for each of the two tasks prior to beginning the vibrotactile conditions in VR. Each baseline trial lasted three minutes. We refrained from expanding the baseline measurements because the trials took approximately two hours to complete and included people with MS who reported reduced physical functioning. To reduce the learning effects, we counterbalanced both tasks and the vibrotactile conditions in VR.

For CoP vibrotactile condition, we streamed the CoPx and CoPy data from the balance board to Unity through sockets in order to provide participants with CoP vibrotactile feedback based on their balance board position. However, it is unknown how lower levels of latency will alter our findings for this scenario. 

One-second intervals were used to provide the "rhythmic" vibrotactile feedback. We did not test this feedback condition across a range of timeframes (e.g., two-second). For this reason, investigations that deliver "rhythmic" vibrotactile stimulation with variable time intervals may result in somewhat different conclusions.

Participants in the static vibrotactile feedback may have become fatigued from listening to the vibration continuously. However, the effect of fatigue in this situation was not quantified.

Noise was produced by the vibrating motors employed for vibrotactile feedback. Thus, it's hard to determine whether the generated noises affected balance improvement or not. To confirm this, further study will be required.

We did not modify the table height for the standing reach and grasp task depending on the participant's height, which may have impacted the findings. However, there were no statistically significant variations in the heights of the participants (Table 1). Therefore, we anticipated a minor impact.

Participants wore harnesses throughout the research to prevent falls, which may have improved their balance somewhat. However, to ensure the consistency and safety of the research protocol, we asked all participants to wear harnesses, regardless of whether they had balance difficulties or not. Consequently, investigations examining balance without a harness may obtain different findings.

The duration of the study was fairly lengthy, and participants were expected to stand while experiencing the vibrotactile feedback. This often led to weariness, and users were required to remove their HMD and relax for a few minutes between sessions. This may have enabled individuals to reestablish spatial awareness and balance, which may have skewed the results.


Due to COVID-19 and our chosen test group, which included individuals with BI caused by MS, the recruiting procedure was challenging since many prospective volunteers had impaired immune systems, putting them at high risk for COVID-19. Therefore, they did not take part in the research. If the research had been conducted outside of COVID-19, more people may have been recruited. In this situation, we could have observed more significant differences across the various study conditions in VR for between-group comparisons. Additional study is necessary to corroborate this.


\section{Conclusion}
In this study, we assessed the influence of several vibrotactile feedback modalities on VR balance. In our investigation, all vibrotactile feedback conditions (spatial, CoP, rhythmic, and static) substantially improved balance in VR. Spatial and CoP vibrotactile feedback outperformed rhythmic and static conditions substantially. There was no statistically significant difference between spatial and CoP conditions, nor between rhythmic and static conditions. Researchers will be better able to comprehend the various types of vibrotactile feedback for maintaining balance in an HMD-based VEs as a consequence of these findings. In addition, this study may assist developers in creating VR experiences that are more accessible and useful for those with and without balance issues. We will investigate locomotion challenges in our future research and evaluate the efficacy of vibrotactile feedback for gait improvement.


\bibliographystyle{abbrv-doi}

\bibliography{template}

\begin{thebibliography}{10}

\bibitem{WinNT}
{[n. d.]} people with disabilities in the world.
\newblock
  \url{https://www.un.org/development/desa/disabilities/resources/factsheet-on-persons-with-disabilities.html}.
\newblock Accessed: 2021-08-30.

\bibitem{agrawal2009disorders}
Y.~Agrawal, J.~P. Carey, C.~C. Della~Santina, M.~C. Schubert, and L.~B. Minor.
\newblock Disorders of balance and vestibular function in us adults: data from
  the national health and nutrition examination survey, 2001-2004.
\newblock {\em Archives of internal medicine}, 169(10):938--944, 2009.

\bibitem{alahakone2010real}
A.~U. Alahakone and S.~A. Senanayake.
\newblock A real-time system with assistive feedback for postural control in
  rehabilitation.
\newblock {\em IEEE/ASME Transactions on Mechatronics}, 15(2):226--233, 2010.

\bibitem{ballardini2020vibrotactile}
G.~Ballardini, V.~Florio, A.~Canessa, G.~Carlini, P.~Morasso, and M.~Casadio.
\newblock Vibrotactile feedback for improving standing balance.
\newblock {\em Frontiers in bioengineering and biotechnology}, 8:94, 2020.

\bibitem{baram2012gait}
Y.~Baram and R.~Lenger.
\newblock Gait improvement in patients with cerebral palsy by visual and
  auditory feedback.
\newblock {\em Neuromodulation: Technology at the Neural Interface},
  15(1):48--52, 2012.

\bibitem{bergeron2015use}
M.~Bergeron, C.~L. Lortie, and M.~J. Guitton.
\newblock Use of virtual reality tools for vestibular disorders rehabilitation:
  a comprehensive analysis.
\newblock {\em Advances in medicine}, 2015, 2015.

\bibitem{bisson2007functional}
E.~Bisson, B.~Contant, H.~Sveistrup, and Y.~Lajoie.
\newblock Functional balance and dual-task reaction times in older adults are
  improved by virtual reality and biofeedback training.
\newblock {\em Cyberpsychology \& behavior}, 10(1):16--23, 2007.

\bibitem{blaszczyk2007assessment}
J.~B{\l}aszczyk, R.~Orawiec, D.~Duda-K{\l}odowska, and G.~Opala.
\newblock Assessment of postural instability in patients with parkinson’s
  disease.
\newblock {\em Experimental Brain Research}, 183(1):107--114, 2007.

\bibitem{bloem1992postural}
B.~Bloem.
\newblock Postural instability in parkinson's disease.
\newblock {\em Clinical neurology and neurosurgery}, 94:41--45, 1992.

\bibitem{bolton2019motor}
D.~A. Bolton, D.~M. Cole, B.~Butler, M.~Mansour, G.~Rydalch, D.~W. McDannald,
  and S.~E. Schwartz.
\newblock Motor preparation for compensatory reach-to-grasp responses when
  viewing a wall-mounted safety handle.
\newblock {\em cortex}, 117:135--146, 2019.

\bibitem{borah2007age}
D.~Borah, S.~Wadhwa, U.~Singh, S.~L. Yadav, M.~Bhattacharjee, and V.~Sindhu.
\newblock Age related changes in postural stability.
\newblock {\em Indian J Physiol Pharmacol}, 51(4):395--404, 2007.

\bibitem{chang2020virtual}
E.~Chang, H.~T. Kim, and B.~Yoo.
\newblock Virtual reality sickness: a review of causes and measurements.
\newblock {\em International Journal of Human--Computer Interaction},
  36(17):1658--1682, 2020.

\bibitem{chiari2005audio}
L.~Chiari, M.~Dozza, A.~Cappello, F.~B. Horak, V.~Macellari, and D.~Giansanti.
\newblock Audio-biofeedback for balance improvement: an accelerometry-based
  system.
\newblock {\em IEEE transactions on biomedical engineering}, 52(12):2108--2111,
  2005.

\bibitem{cho2016treadmill}
C.~Cho, W.~Hwang, S.~Hwang, and Y.~Chung.
\newblock Treadmill training with virtual reality improves gait, balance, and
  muscle strength in children with cerebral palsy.
\newblock {\em The Tohoku journal of experimental medicine}, 238(3):213--218,
  2016.

\bibitem{chong2020audio}
U.~Chong and S.~Alimardanov.
\newblock Audio augmented reality using unity for marine tourism.
\newblock In {\em International Conference on Intelligent Human Computer
  Interaction}, pp. 303--311. Springer, 2020.

\bibitem{cordova2014older}
A.~Cordova and C.~Gabbard.
\newblock Do older adults perceive postural constraints for reach estimation?
\newblock {\em Experimental aging research}, 40(5):578--588, 2014.

\bibitem{coren1993lateral}
S.~Coren.
\newblock The lateral preference inventory for measurement of handedness,
  footedness, eyedness, and earedness: Norms for young adults.
\newblock {\em Bulletin of the Psychonomic Society}, 31(1):1--3, 1993.

\bibitem{de2016effect}
I.~J. De~Rooij, I.~G. Van De~Port, and J.-W.~G. Meijer.
\newblock Effect of virtual reality training on balance and gait ability in
  patients with stroke: systematic review and meta-analysis.
\newblock {\em Physical therapy}, 96(12):1905--1918, 2016.

\bibitem{denomme2014understanding}
L.~T. Denomm{\'e}, P.~Mandalfino, and M.~E. Cinelli.
\newblock Understanding balance differences in individuals with multiple
  sclerosis with mild disability: an investigation of differences in sensory
  feedback on postural control during a romberg task.
\newblock {\em Experimental brain research}, 232(6):1833--1842, 2014.

\bibitem{diener1988role}
H.-C. Diener and J.~Dichgans.
\newblock On the role of vestibular, visual and somatosensory information for
  dynamic postural control in humans.
\newblock {\em Progress in brain research}, 76:253--262, 1988.

\bibitem{duque2013effects}
G.~Duque, D.~Boersma, G.~Loza-Diaz, S.~Hassan, H.~Suarez, D.~Geisinger,
  P.~Suriyaarachchi, A.~Sharma, and O.~Demontiero.
\newblock Effects of balance training using a virtual-reality system in older
  fallers.
\newblock {\em Clinical interventions in aging}, 8:257, 2013.

\bibitem{epure2014effect}
P.~Epure, C.~Gheorghe, T.~Nissen, L.-O. Toader, A.~Nicolae, S.~S. Nielsen,
  D.~J.~R. Christensen, A.~L. Brooks, and E.~Petersson.
\newblock Effect of the oculus rift head mounted display on postural stability.
\newblock In {\em The 10th International Conference on Disability Virtual
  Reality \& Associated Technologies: Proceedings}, pp. 119--127. Reading
  University Press, 2014.

\bibitem{ferdous2016visual}
S.~M.~S. Ferdous, I.~M. Arafat, and J.~Quarles.
\newblock Visual feedback to improve the accessibility of head-mounted displays
  for persons with balance impairments.
\newblock In {\em 2016 IEEE Symposium on 3D User Interfaces (3DUI)}, pp.
  121--128. IEEE, 2016.

\bibitem{ferdous2018investigating}
S.~M.~S. Ferdous, T.~I. Chowdhury, I.~M. Arafat, and J.~Quarles.
\newblock Investigating the reason for increased postural instability in
  virtual reality for persons with balance impairments.
\newblock In {\em Proceedings of the 24th ACM Symposium on Virtual Reality
  Software and Technology}, pp. 1--7, 2018.

\bibitem{franco2012ibalance}
C.~Franco, A.~Fleury, P.-Y. Gum{\'e}ry, B.~Diot, J.~Demongeot, and
  N.~Vuillerme.
\newblock ibalance-abf: a smartphone-based audio-biofeedback balance system.
\newblock {\em IEEE transactions on biomedical engineering}, 60(1):211--215,
  2012.

\bibitem{ghai2018effect}
S.~Ghai, I.~Ghai, and A.~O. Effenberg.
\newblock Effect of rhythmic auditory cueing on aging gait: a systematic review
  and meta-analysis.
\newblock {\em Aging and disease}, 9(5):901, 2018.

\bibitem{goodworth2009influence}
A.~D. Goodworth, C.~Wall~III, and R.~J. Peterka.
\newblock Influence of feedback parameters on performance of a vibrotactile
  balance prosthesis.
\newblock {\em IEEE Transactions on Neural Systems and Rehabilitation
  Engineering}, 17(4):397--408, 2009.

\bibitem{guo2013effects}
R.~Guo, G.~Samaraweera, and J.~Quarles.
\newblock The effects of ves on mobility impaired users: Presence, gait, and
  physiological response.
\newblock In {\em Proceedings of the 19th ACM Symposium on Virtual Reality
  Software and Technology}, pp. 59--68, 2013.

\bibitem{guo2015mobility}
R.~Guo, G.~Samaraweera, and J.~Quarles.
\newblock Mobility impaired users respond differently than healthy users in
  virtual environments.
\newblock {\em Computer Animation and Virtual Worlds}, 26(5):509--526, 2015.

\bibitem{hasegawa2017learning}
N.~Hasegawa, K.~Takeda, M.~Sakuma, H.~Mani, H.~Maejima, and T.~Asaka.
\newblock Learning effects of dynamic postural control by auditory biofeedback
  versus visual biofeedback training.
\newblock {\em Gait \& posture}, 58:188--193, 2017.

\bibitem{helps2014different}
S.~K. Helps, S.~Bamford, E.~J. Sonuga-Barke, and G.~B. S{\"o}derlund.
\newblock Different effects of adding white noise on cognitive performance of
  sub-, normal and super-attentive school children.
\newblock {\em PloS one}, 9(11):e112768, 2014.

\bibitem{henry2019age}
M.~Henry and S.~Baudry.
\newblock Age-related changes in leg proprioception: implications for postural
  control.
\newblock {\em Journal of neurophysiology}, 122(2):525--538, 2019.

\bibitem{huang2015effects}
M.~H. Huang and S.~H. Brown.
\newblock Effects of task context during standing reach on postural control in
  young and older adults: A pilot study.
\newblock {\em Gait \& posture}, 41(1):276--281, 2015.

\bibitem{kennedy1993simulator}
R.~S. Kennedy, N.~E. Lane, K.~S. Berbaum, and M.~G. Lilienthal.
\newblock Simulator sickness questionnaire: An enhanced method for quantifying
  simulator sickness.
\newblock {\em The international journal of aviation psychology},
  3(3):203--220, 1993.

\bibitem{kido2010postural}
T.~Kido, Y.~Tabara, M.~Igase, N.~Ochi, E.~Uetani, T.~Nagai, M.~Yamamoto,
  K.~Taguchi, T.~Miki, and K.~Kohara.
\newblock Postural instability is associated with brain atrophy and cognitive
  impairment in the elderly: the j-shipp study.
\newblock {\em Dementia and geriatric cognitive disorders}, 29(5):379--387,
  2010.

\bibitem{kim2021clinical}
H.~Kim, D.~J. Kim, W.~H. Chung, K.-A. Park, J.~D. Kim, D.~Kim, K.~Kim, and
  H.~J. Jeon.
\newblock Clinical predictors of cybersickness in virtual reality (vr) among
  highly stressed people.
\newblock {\em Scientific reports}, 11(1):1--11, 2021.

\bibitem{kingma2019vibrotactile}
H.~Kingma, L.~Felipe, M.-C. Gerards, P.~Gerits, N.~Guinand, A.~Perez-Fornos,
  V.~Demkin, and R.~Van De~Berg.
\newblock Vibrotactile feedback improves balance and mobility in patients with
  severe bilateral vestibular loss.
\newblock {\em Journal of neurology}, 266(1):19--26, 2019.

\bibitem{li2016reliability}
Z.~Li, Y.-Y. Liang, L.~Wang, J.~Sheng, and S.-J. Ma.
\newblock Reliability and validity of center of pressure measures for balance
  assessment in older adults.
\newblock {\em Journal of physical therapy science}, 28(4):1364--1367, 2016.

\bibitem{lott2003effect}
A.~Lott, E.~Bisson, Y.~Lajoie, J.~McComas, and H.~Sveistrup.
\newblock The effect of two types of virtual reality on voluntary center of
  pressure displacement.
\newblock {\em Cyberpsychology \& behavior}, 6(5):477--485, 2003.

\bibitem{martinez2018analysing}
A.~Martinez, A.~I. Paganelli, and A.~Raposo.
\newblock Analysing balance loss in vr interaction with hmds.
\newblock {\em Journal on Interactive Systems}, 9(2), 2018.

\bibitem{meldrum2015effectiveness}
D.~Meldrum, S.~Herdman, R.~Vance, D.~Murray, K.~Malone, D.~Duffy, A.~Glennon,
  and R.~McConn-Walsh.
\newblock Effectiveness of conventional versus virtual reality--based balance
  exercises in vestibular rehabilitation for unilateral peripheral vestibular
  loss: results of a randomized controlled trial.
\newblock {\em Archives of physical medicine and rehabilitation},
  96(7):1319--1328, 2015.

\bibitem{niehorster2017accuracy}
D.~C. Niehorster, L.~Li, and M.~Lappe.
\newblock The accuracy and precision of position and orientation tracking in
  the htc vive virtual reality system for scientific research.
\newblock {\em i-Perception}, 8(3):2041669517708205, 2017.

\bibitem{park2015effects}
E.-C. Park, S.-G. Kim, and C.-W. Lee.
\newblock The effects of virtual reality game exercise on balance and gait of
  the elderly.
\newblock {\em Journal of physical therapy science}, 27(4):1157--1159, 2015.

\bibitem{pinkl2020spatialized}
J.~Pinkl and M.~Cohen.
\newblock Spatialized ar polyrhythmic metronome using bose frames eyewear.
\newblock 2020.

\bibitem{powell1995activities}
L.~E. Powell and A.~M. Myers.
\newblock The activities-specific balance confidence (abc) scale.
\newblock {\em The Journals of Gerontology Series A: Biological Sciences and
  Medical Sciences}, 50(1):M28--M34, 1995.

\bibitem{rendon2012effect}
A.~A. Rendon, E.~B. Lohman, D.~Thorpe, E.~G. Johnson, E.~Medina, and
  B.~Bradley.
\newblock The effect of virtual reality gaming on dynamic balance in older
  adults.
\newblock {\em Age and ageing}, 41(4):549--552, 2012.

\bibitem{robert2016effect}
M.~T. Robert, L.~Ballaz, and M.~Lemay.
\newblock The effect of viewing a virtual environment through a head-mounted
  display on balance.
\newblock {\em Gait \& posture}, 48:261--266, 2016.

\bibitem{ross2016auditory}
J.~Ross, O.~Will, Z.~McGann, and R.~Balasubramaniam.
\newblock Auditory white noise reduces age-related fluctuations in balance.
\newblock {\em Neuroscience letters}, 630:216--221, 2016.

\bibitem{ruhe2011center}
A.~Ruhe, R.~Fejer, and B.~Walker.
\newblock Center of pressure excursion as a measure of balance performance in
  patients with non-specific low back pain compared to healthy controls: a
  systematic review of the literature.
\newblock {\em European Spine Journal}, 20(3):358--368, 2011.

\bibitem{rust2020benefits}
H.~Rust, N.~Lutz, V.~Zumbrunnen, M.~Imhof, {\"O}.~Yaldizli, V.~Haller, and
  J.~H. Allum.
\newblock Benefits of short-term training with vibrotactile biofeedback of
  trunk sway on balance control in multiple sclerosis.
\newblock {\em Physical Medicine and Rehabilitation Research}, 5(1):1--10,
  2020.

\bibitem{samaraweera2013latency}
G.~Samaraweera, R.~Guo, and J.~Quarles.
\newblock Latency and avatars in virtual environments and the effects on gait
  for persons with mobility impairments.
\newblock In {\em 2013 IEEE Symposium on 3D User Interfaces (3DUI)}, pp.
  23--30. IEEE, 2013.

\bibitem{schepens2010short}
S.~Schepens, A.~Goldberg, and M.~Wallace.
\newblock The short version of the activities-specific balance confidence (abc)
  scale: its validity, reliability, and relationship to balance impairment and
  falls in older adults.
\newblock {\em Archives of gerontology and geriatrics}, 51(1):9--12, 2010.

\bibitem{sienko2017role}
K.~Sienko, S.~Whitney, W.~Carender, and C.~Wall~III.
\newblock The role of sensory augmentation for people with vestibular deficits:
  real-time balance aid and/or rehabilitation device?
\newblock {\em Journal of Vestibular Research}, 27(1):63--76, 2017.

\bibitem{sienko2012biofeedback}
K.~H. Sienko, M.~D. Balkwill, and C.~Wall.
\newblock Biofeedback improves postural control recovery from multi-axis
  discrete perturbations.
\newblock {\em Journal of neuroengineering and rehabilitation}, 9(1):1--11,
  2012.

\bibitem{sienko2018potential}
K.~H. Sienko, R.~D. Seidler, W.~J. Carender, A.~D. Goodworth, S.~L. Whitney,
  and R.~J. Peterka.
\newblock Potential mechanisms of sensory augmentation systems on human balance
  control.
\newblock {\em Frontiers in neurology}, 9:944, 2018.

\bibitem{soffel2016postural}
F.~Soffel, M.~Zank, and A.~Kunz.
\newblock Postural stability analysis in virtual reality using the htc vive.
\newblock In {\em Proceedings of the 22nd ACM Conference on Virtual Reality
  Software and Technology}, pp. 351--352, 2016.

\bibitem{soltani2020influence}
P.~Soltani and R.~Andrade.
\newblock The influence of virtual reality head-mounted displays on balance
  outcomes and training paradigms: A systematic review.
\newblock {\em Frontiers in sports and active living}, 2:233, 2020.

\bibitem{takahashi2001change}
Y.~Takahashi and A.~Murata.
\newblock Change of equilibrium under the influence of vr experience.
\newblock In {\em Proceedings 10th IEEE International Workshop on Robot and
  Human Interactive Communication. ROMAN 2001 (Cat. No. 01TH8591)}, pp.
  642--647. IEEE, 2001.

\bibitem{tan2012anticipatory}
C.~Tan, J.~Tretriluxana, E.~Pitsch, N.~Runnarong, and C.~J. Winstein.
\newblock Anticipatory planning of functional reach-to-grasp: a pilot study.
\newblock {\em Neurorehabilitation and neural repair}, 26(8):957--967, 2012.

\bibitem{thompson2017balance}
L.~A. Thompson, M.~Badache, S.~Cale, L.~Behera, and N.~Zhang.
\newblock Balance performance as observed by center-of-pressure parameter
  characteristics in male soccer athletes and non-athletes.
\newblock {\em Sports}, 5(4):86, 2017.

\bibitem{velazquez2010wearable}
R.~Vel{\'a}zquez.
\newblock Wearable assistive devices for the blind.
\newblock In {\em Wearable and autonomous biomedical devices and systems for
  smart environment}, pp. 331--349. Springer, 2010.

\bibitem{wall2009vibrotactile}
C.~Wall~III, D.~M. Wrisley, and K.~D. Statler.
\newblock Vibrotactile tilt feedback improves dynamic gait index: a fall risk
  indicator in older adults.
\newblock {\em Gait \& posture}, 30(1):16--21, 2009.

\bibitem{wannstedt1978use}
G.~T. Wannstedt and R.~M. Herman.
\newblock Use of augmented sensory feedback to achieve symmetrical standing.
\newblock {\em Physical Therapy}, 58(5):553--559, 1978.

\bibitem{young2011assessing}
W.~Young, S.~Ferguson, S.~Brault, and C.~Craig.
\newblock Assessing and training standing balance in older adults: a novel
  approach using the ‘nintendo wii’balance board.
\newblock {\em Gait \& posture}, 33(2):303--305, 2011.

\end{thebibliography}
\end{document}